\documentclass[conference, twocolumn]{IEEEtran}

\usepackage[usenames,dvipsnames]{xcolor}
\usepackage{amsmath,amssymb,amsfonts,tikz,pgfplots,bm}
\usepackage{cite,booktabs,tabularx}
\usepackage{algorithmic}
\usepackage{graphicx}
\usepackage{textcomp}
\usepackage{verbatim}
\usepackage{amsmath}
\usepackage[ruled,vlined]{algorithm2e}
\usepackage{nicefrac}
\usepackage[inline,shortlabels]{enumitem}
\usepackage{array}
\usepackage{balance}
\usepackage{multirow}

\usepackage[compact]{titlesec}         %
\titlespacing{\subsection}{5pt}{8pt}{4pt}%

\newcommand{\cN}{{\cal N}}
\newcommand{\cO}{{\cal O}}

\newcommand{\cS}{{\cal S}}

\newcommand{\bfd}{{\boldsymbol d}}

\newcommand{\bfq}{{\boldsymbol q}}
\newcommand{\bfr}{{\boldsymbol r}}

\newcommand{\bfu}{{\boldsymbol u}}
\newcommand{\bfv}{{\boldsymbol v}}
\newcommand{\bfw}{{\boldsymbol w}}
\newcommand{\bfx}{{\boldsymbol x}}
\newcommand{\bfy}{{\boldsymbol y}}
\newcommand{\bfz}{{\boldsymbol z}}

\newcommand{\bfC}{{\boldsymbol C}}
\newcommand{\bfD}{{\boldsymbol D}}

\newcommand{\bfX}{{\boldsymbol X}}
\newcommand{\bfY}{{\boldsymbol Y}}
\newcommand{\bfZ}{{\boldsymbol Z}}

\usetikzlibrary{positioning,calc}
\tikzstyle{block} = [rectangle, draw, text centered, rounded corners, minimum height=1.5em,fill=black!5!]

\def\approxprop{%
	\def\p{%
		\setbox0=\vbox{\hbox{$\propto$}}%
		\ht0=0.6ex \box0 }%
	\def\s{%
		\vbox{\hbox{$\sim$}}%
	}%
	\mathrel{\raisebox{0.7ex}{%
			\mbox{$\underset{\s}{\p}$}%
	}}%
}

\newcommand{\nums}{\ensuremath{M}}

\newcommand{\Rtot}{\ensuremath{R}}
\newcommand{\Rout}{\ensuremath{R_\mathrm{o}}}
\newcommand{\Rin}{\ensuremath{R_\mathrm{i}}}
\newcommand{\Rix}{\ensuremath{R_\mathrm{ix}}}

\newcommand{\outY}{\ensuremath{\bfY}}
\newcommand{\outC}{\ensuremath{\bfC}}
\newcommand{\outS}{\ensuremath{\cS}}
\newcommand{\listset}{list}

\newcommand{\nix}{\ensuremath{n_{\mathsf{ix}}}}
\newcommand{\kix}{\ensuremath{k_{\mathsf{ix}}}}

\newcommand{\nin}{\ensuremath{n_{\mathsf{i}}}}
\newcommand{\kin}{\ensuremath{k_{\mathsf{i}}}}

\newcommand{\chq}{\ensuremath{4}}

\newcommand{\Rbcjr}{\ensuremath{R_{\text{BCJR-once}}}}

\newcommand{\rbcjr}{\ensuremath{r_{\text{BCJR-once}}^{\bfd}}}

\newcommand{\lseq}{\ensuremath{L}}%
\newcommand{\Sig}{\ensuremath{\Sigma}}

\newcommand{\mi}{\ensuremath{\mathrm{I}}}

\newcommand{\totdraw}{\ensuremath{N}}
\newcommand{\prd}{\ensuremath{p_{\mathrm{D}}}}
\newcommand{\pri}{\ensuremath{p_{\mathrm{I}}}}
\newcommand{\prs}{\ensuremath{p_{\mathrm{S}}}}
\newcommand{\prt}{\ensuremath{p_{\mathrm{T}}}}
\newcommand{\pout}{\ensuremath{p_{\mathrm{out}}}}
\newcommand{\qout}{\ensuremath{q_{\mathrm{out}}}}

\newcommand{\scmem}{\ensuremath{m}}

\newcommand{\leno}{\ensuremath{n_{\mathsf{o}}}}
\newcommand{\bleno}{\ensuremath{L_{\mathsf{o}}}}

\newcommand{\dimo}{\ensuremath{k_{\mathsf{o}}}}
\newcommand{\field}{\ensuremath{\mathbb{F}}}
\newcommand{\outq}{\ensuremath{4}}

\newcommand{\numc}{\ensuremath{\nums'}}

\newcommand{\llr}{\ensuremath{\mathrm{LLR}}}

\newcommand{\eucd}[1]{\ensuremath{d_{\mathrm{E}}^{(#1)}}}
\newcommand{\binL}{\ensuremath{L_{\mathrm{B}}}}

\newcommand{\bfwi}[1]{\ensuremath{\bfw^{(#1)}}}
\newcommand{\bfvi}[1]{\ensuremath{\bfv^{(#1)}}}
\newcommand{\bfii}[1]{\ensuremath{\mathrm{ind}(#1)}}
\newcommand{\bfrj}[1]{\ensuremath{\bfr^{(#1)}}}
\newcommand{\wit}{\ensuremath{w_t^{(i)}}}

\IEEEoverridecommandlockouts

\begin{document}

\title{Index-Based Concatenated Codes\\for the Multi-Draw DNA Storage Channel}

\author{\IEEEauthorblockN{{\bf Lorenz Welter}\IEEEauthorrefmark{1}, {\bf Issam Maarouf}\IEEEauthorrefmark{2}, {\bf Andreas Lenz}\IEEEauthorrefmark{1},\\ {\bf Antonia Wachter-Zeh}\IEEEauthorrefmark{1}, {\bf Eirik Rosnes}\IEEEauthorrefmark{2}, {and \textbf{Alexandre Graell i Amat}}\IEEEauthorrefmark{3}}

\IEEEauthorblockA{
	\IEEEauthorrefmark{1}School of Computation, Information and Technology, Technical University of Munich, DE-80333 Munich, Germany
}

\IEEEauthorblockA{
	\IEEEauthorrefmark{2}Simula UiB, N-5006  Bergen,  Norway
}

\IEEEauthorblockA{
	\IEEEauthorrefmark{3}Department of Electrical Engineering, Chalmers University of Technology, SE-41296 Gothenburg, Sweden
}%

\thanks{
The work of L. Welter, A. Lenz, and A. Wachter-Zeh has been supported by the European Research
Council (ERC) under the European Union’s Horizon 2020 research and innovation programme (Grant Agreement No. 801434).

The work of A. Graell i Amat was supported by the Swedish Research Council under grant 2020-03687.} \vspace{-2.5em}}

\maketitle

\begin{abstract}
We consider error-correcting coding for DNA-based storage. We  model the DNA storage channel as a \emph{multi-draw IDS channel} where the input data is chunked into $\nums$ short DNA strands, which are copied a random number of times, and the channel outputs a random selection of $\totdraw$ noisy DNA strands. The retrieved DNA strands are prone to insertion, deletion, and substitution (IDS) errors. We propose an index-based concatenated coding scheme consisting of the concatenation of an outer code, an index code, and an inner \emph{synchronization} code, where the latter two tackle IDS errors. We further propose a mismatched joint index-synchronization code maximum  a posteriori probability decoder with optional clustering to infer symbolwise a posteriori probabilities for the outer decoder. We compute achievable information rates for the outer code and present Monte-Carlo simulations for information-outage probabilities and frame error rates on synthetic and experimental data, respectively.
\end{abstract}

\IEEEpeerreviewmaketitle

\section{Introduction} \label{sec:introduction}

This work %
aims at improving the reliability of DNA-based data storage. 
We analyze the \emph{multi-draw IDS channel} which is an abstraction of
the synthesis (writing), storage, and sequencing (reading) procedures, including insertion, deletion, and substitution (IDS) errors,  thereby narrowing the modeling gap to the real DNA storage channel~\cite{heckel_characterization_2019}. 

The capacity of the DNA storage channel
has been studied in \cite{heckel_fundamental_ISIT_2017,lenz_upperboundcapDNA_2019, lenz_achieving_2020,weinberger_dna_2022}; see \cite{HeckelShomorony_Foundations_2022} for an  overview. 
For coping with IDS errors in a single-sequence transmission, the classical scheme in \cite{davey_reliable_2001} with the improved decoding method from \cite{briffa_improved_2010} is most relevant for our work. %
Independently, \cite{Srinivasavaradhan2021TrellisBMA, Maarouf2022ConcatenatedCF} analyzed coded trace reconstruction under IDS errors and
showed significant gains by leveraging multiple noisy copies of the same transmit sequence even with sub-optimal decoding techniques. 

Inspired by experimental works \cite{church_next-generation_2012,goldman_towards_2013,grass_robust_2015,yazdi_rewritable_2015,organick_random_2018,antkowiak_low_2020}, we propose an index-based concatenated coding scheme for the multi-draw IDS channel. %
Index-based schemes for the multi-draw IDS channel are sub-optimal, albeit  very practical \cite{HeckelShomorony_Foundations_2022,lenz_coding_2020-1}.
Similar coding approaches are analyzed in \cite{lenz_concatachieve_2020,weinberger_codedindex_2022}. In our scheme, the information is encoded by an outer code to provide overall error protection, primarily coping with unresolved errors and strand erasures of the data. The resulting codeword is split into $\nums$ data blocks. In each data block, index information is embedded after being encoded by a low-rate index code, whereas the data block itself is encoded by an inner code. Both codes are tailored to deal with IDS errors. The $\nums$ resulting strands are then transmitted over the multi-draw IDS channel, which models the DNA storage channel. At the receiver, each noisy strand is decoded separately by a joint index and inner maximum a posteriori probability (MAP) decoder outputting symbolwise a posteriori probabilities (APPs). 
To leverage the multi-copy gain, we combine the APPs according to the subsequent clustering and index decoding. We optionally cluster the received strands based on the obtained APPs, benefiting from the error-correction capabilities of the index and inner codes. %
For each cluster, an index decision is made by jointly considering the received strands in a cluster. The APPs are ordered according to the index decisions and fed into a soft-input outer decoder. 

We analyze our proposed scheme in terms of achievable information rates (AIRs). The AIRs provide insights to the performance of different index-inner coding and decoding techniques and  a benchmark for an outer code. For different coding setups in the finite blocklength regime, we compute information-outage probabilities on synthetic data and present frame error rates (FERs) on experimental data from \cite{Srinivasavaradhan2021TrellisBMA}.

\section{DNA Storage Channel Model} \label{sec:system}

\subsection{IDS Channel}
We consider a model in which IDS errors are independent and identically distributed. Let $\bfx=(x_1,\ldots, x_{\lseq})$ and $\bfy=(y_1,\ldots,y_{\lseq'})$ be the DNA strand to be synthesized and a single read at the output of the sequencing process, respectively, with $x_t,y_t\in \Sig_\chq=\{\mathsf{A},\mathsf{C},\mathsf{G},\mathsf{T}\}$.%

The input to the channel can be seen as a queue in which  symbols $x_t$ are successively enqueued for transmission. The output strand $\bfy$ is generated as follows:
$\bfy$ is first initialized to an empty vector before  $x_1$ is enqueued. Then, for each input symbol $x_t$, the following three events may occur: 
\begin{enumerate*}
    \item A random symbol $a \in \Sig _\chq$ is inserted with probability $\pri$. In this case, $\bfy$ is concatenated with symbol $a$ as $\bfy \leftarrow (\bfy , a)$ and $x_t$ remains in the queue.
    \item The symbol $x_t$ is deleted with probability $\prd$, $\bfy$ remains unchanged, and $x_{t+1}$ is enqueued.
    \item The symbol $x_t$ is transmitted with probability $\prt=1-\pri-\prd$. In this case, the symbol is received correctly with probability $1-\prs$ or incorrectly with probability $\prs$, in which case the symbol is
       substituted by a symbol $a' \in \Sig_\chq \setminus \{x_t\}$ picked uniformly at random. The output is set to $\bfy \leftarrow (\bfy , x_t)$ and $\bfy \leftarrow (\bfy , a')$, respectively, and $x_{t+1}$ is enqueued.
\end{enumerate*}
After  $x_{\lseq}$ leaves the queue, we obtain the output strand $\bfy$, of length $\lseq'$. Note that $\lseq^\prime$ is random and depends on the channel realization. Further, the symbols in $\bfy$ are not synchronized anymore. Hence, $y_t$ could be the result of transmitting a symbol $x_{t^\prime}$ with $t^\prime \neq t$.

\subsection{Multi-Draw IDS Channel}\label{sec:Mult-draw}
The data sequence is divided into $M$ short DNA strands $\bfx_1,\ldots,\bfx_M$, each of length $\lseq$, which are synthesized and stored in a \emph{pool}. 
We define the input list $\bfX \triangleq ( \bfx_1, \dots, \bfx_\nums )$. We assume that the PCR amplification and  sequencing processes generate $N=cM$ reads from the strands in the pool, resulting in a random number of (noisy) reads for each strand $\bfx_i$. The factor $c$ is a positive real number referred to as the \emph{coverage depth} in the literature. 

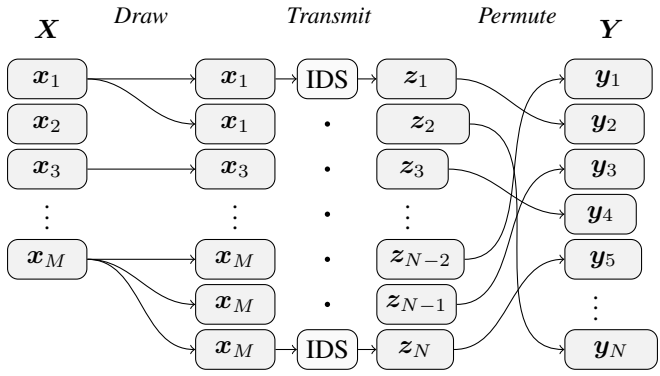
\begin{figure}[t]
	\centering
				\begin{tikzpicture}
			
			\def\xydist{0.6};
			\def\xxyydist{2.5};
			\def\yydist{0.6};
			
			\def\bwidth{3.0em};

			\tikzstyle{seqblock} = [rectangle, draw, text centered, rounded corners, minimum height=1.5em,fill=black!5!, minimum width=\bwidth]
			\tikzset{
				zblock/.style = {
					minimum width= {#1}em
					},
				rectangle, draw, text centered, rounded corners, minimum height=1.5em,fill=black!5!
			}
			\tikzstyle{idsblock} = [rectangle, draw, text centered, rounded corners, minimum height=1.5em,fill=black!2!, minimum width=0.5*\bwidth]
			
			\foreach \ii in {1,...,3}
				{\node[seqblock] (x\ii) at (0,-1*\ii*\xydist) {$\bfx_\ii$};	}
				
			\node (x4) at (0,-1*3.9*\xydist) {$\vdots$};
				
			\node[seqblock] (x5) at (0,-1*5*\xydist) {$\bfx_\nums$};

            \foreach \ii in {1,...,2}
				{\node[seqblock] (x1\ii) at (1*\xxyydist,-1*\ii*\xydist) {$\bfx_1$};	}

            \foreach \ii in {1}
				{\node[seqblock] (x2\ii) at (1*\xxyydist,-2*\xydist-1*\ii*\xydist) {$\bfx_3$};	}
    
			\node (x4) at (1*\xxyydist,-3.9*\xydist) {$\vdots$};

            \foreach \ii in {1,...,3}
				{\node[seqblock] (x5\ii) at (1*\xxyydist,-4*\xydist-1*\ii*\xydist) {$\bfx_\nums$};	}

			\node[rectangle, draw, text centered, rounded corners, minimum height=1.5em,fill=black!5!, minimum width=\bwidth, anchor=west] (z1) at (1.75*\xxyydist,-1*1*\yydist) {$\bfz_1$};
			\node[rectangle, draw, text centered, rounded corners, minimum height=1.5em,fill=black!5!, minimum width=\bwidth+0.5em, anchor=west] (z2) at (1.75*\xxyydist,-1*2*\yydist) {$\bfz_2$};
			\node[rectangle, draw, text centered, rounded corners, minimum height=1.5em,fill=black!5!, minimum width=\bwidth-0.3em, anchor=west] (z3) at (1.75*\xxyydist,-1*3*\yydist) {$\bfz_3$};
			\node[rectangle, draw, text centered, rounded corners, minimum height=1.5em,fill=black!5!, minimum width=\bwidth+0.3em, anchor=west] (z4) at (1.75*\xxyydist,-1*5*\yydist) {$\bfz_{\totdraw-2}$};
			\node[rectangle, draw, text centered, rounded corners, minimum height=1.5em,fill=black!5!, minimum width=\bwidth+0.0em, anchor=west] (z5) at (1.75*\xxyydist,-1*6*\yydist) {$\bfz_{\totdraw-1}$};
			\node[anchor=west] (z6) at (1.85*\xxyydist,-1*3.9*\yydist) {$\vdots$};
			\node[rectangle, draw, text centered, rounded corners, minimum height=1.5em,fill=black!5!, minimum width=\bwidth-0.1em, anchor=west] (z7) at (1.75*\xxyydist,-1*7*\yydist) {$\bfz_\totdraw$};

            \node[rectangle, draw, text centered, rounded corners, minimum height=1.5em,fill=black!5!, minimum width=\bwidth+0.3em, anchor=west] (y1) at (2.75*\xxyydist,-1*1*\yydist) {$\bfy_1$};
			\node[rectangle, draw, text centered, rounded corners, minimum height=1.5em,fill=black!5!, minimum width=\bwidth, anchor=west] (y2) at (2.75*\xxyydist,-1*2*\yydist) {$\bfy_2$};
			\node[rectangle, draw, text centered, rounded corners, minimum height=1.5em,fill=black!5!, minimum width=\bwidth+0.0em, anchor=west] (y3) at (2.75*\xxyydist,-1*3*\yydist) {$\bfy_3$};
			\node[rectangle, draw, text centered, rounded corners, minimum height=1.5em,fill=black!5!, minimum width=\bwidth-0.3em, anchor=west] (y4) at (2.75*\xxyydist,-1*4*\yydist) {$\bfy_4$};
			\node[rectangle, draw, text centered, rounded corners, minimum height=1.5em,fill=black!5!, minimum width=\bwidth-0.1em, anchor=west] (y5) at (2.75*\xxyydist,-1*5*\yydist) {$\bfy_5$};
			\node[anchor=west] (z6) at (2.85*\xxyydist,-1*5.9*\yydist) {$\vdots$};
			\node[rectangle, draw, text centered, rounded corners, minimum height=1.5em,fill=black!5!, minimum width=\bwidth+0.5em, anchor=west] (y7) at (2.75*\xxyydist,-1*7*\yydist) {$\bfy_\totdraw$};

            \node[idsblock] (ids1) at (1.49*\xxyydist,-1*1*\yydist) {IDS}; %
			\draw[->] (x11.east) -- (ids1.west) ;
			\draw[->] (ids1.east) -- (z1.west) ;

               \node[idsblock] (ids2) at (1.49*\xxyydist,-1*7*\yydist) {IDS}; %
			\draw[->] (x53.east) -- (ids2.west) ;
			\draw[->] (ids2.east) -- (z7.west) ;
					
			\foreach \ii in {2,...,6}
			{
				\node[circle, fill=black, minimum size=2pt,
				inner sep=0pt, outer sep=0pt] at (1.49*\xxyydist,-1*\ii*\yydist) {};
			}

            \draw[->] (x1.east) to[out=0,in=-180] node[midway,right,inner sep=2pt] {} (x11.west);
			\draw[->] (x1.east) to[out=0,in=-180] node[midway,right,inner sep=2pt] {} (x12.west);
			\draw[->] (x3.east) to[out=0,in=-180] node[midway,right,inner sep=2pt] {} (x21.west);
			\draw[->] (x5.east) to[out=0,in=-180] node[midway,right,inner sep=2pt] {} (x51.west);
			\draw[->] (x5.east) to[out=0,in=-180] node[midway,right,inner sep=2pt] {} (x52.west);
			\draw[->] (x5.east) to[out=0,in=-180] node[midway,right,inner sep=2pt] {} (x53.west);
		
			\draw[->] (z1.east) to[out=0,in=-180] node[midway,right,inner sep=2pt] {} (y2.west);
			\draw[->] (z2.east) to[out=0,in=-180] node[midway,right,inner sep=2pt] {} (y7.west);
			\draw[->] (z3.east) to[out=0,in=-180] node[midway,right,inner sep=2pt] {} (y4.west);
			\draw[->] (z4.east) to[out=0,in=-180] node[midway,right,inner sep=2pt] {} (y1.west);
			\draw[->] (z5.east) to[out=0,in=-180] node[midway,right,inner sep=2pt] {} (y3.west);
			\draw[->] (z7.east) to[out=0,in=-180] node[midway,right,inner sep=2pt] {} (y5.west);

			\node (x0) at (0,0.1*\xydist) {$\bfX$};			
			\node (z0) at (3.0*\xxyydist,0.1*\xydist) {$\outY$};
			\node (samp) at (0.5*\xxyydist,0.4*\yydist) {\small \textit{Draw}};
             \node (perm) at (2.5*\xxyydist,0.4*\yydist) {\small \textit{Permute}};
			\node (trans) at (1.5*\xxyydist,0.4*\yydist) {\small \textit{Transmit}};
			\end{tikzpicture}
	\caption{Multi-draw IDS channel.}
	\label{fig:multi-draw-channel}
 \vspace{-3.5ex}
\end{figure}

The channel between the input to be synthesized, $\bfX$, and the output of the sequencing process, $\outY$, can be modeled in three phases as explained in the following.
\begin{enumerate}

     \item \emph{Draw:} $\totdraw$ draws are performed from the list $\bfX$ uniformly at random with replacement. Let $D_i$ be the random variable (RV) corresponding to the number of times strand $\bfx_i$ is drawn and define the random vector $\bfD \triangleq (D_1,\dots,D_\nums)$, with $\sum_i D_i = N$. 
    Then, the random vector $\bfD = (D_1,\dots,D_\nums)$ is multinomially distributed as $\bfD \sim \mathsf{Multinom}(\nums, \totdraw, 1/\nums)$. 

     \item \emph{Transmit:} The drawn strands are transmitted through independent IDS channels with identical parameters $\pri$, $\prd$, and $\prs$. We denote by $\bfz_j$ the output of the $j$-th IDS channel, $j\in \{1,\dots,\totdraw\}$, and define $\bfZ \triangleq (\bfz_1,\dots,\bfz_N)$.

     \item \emph{Permute:} The final output of the channel, denoted as the \listset\ $\outY \triangleq (\bfy_1,\dots, \bfy_\totdraw )$ is obtained by a  permutation of $\bfZ$ chosen uniformly at random.
\end{enumerate}
We illustrate the channel model in Fig.~\ref{fig:multi-draw-channel}.\footnote{In  \cite{weinberger_dna_2022, lenz_upperboundcapDNA_2019, HeckelShomorony_Foundations_2022}, the \emph{draw} and \emph{permute} steps are considered as a joint process referred to as \emph{sampling}. We split the step for the sake of presentation; our channel model is essentially the same channel model, only that we consider the noise channel to be an IDS channel.} We define the channel parameter $\beta = \frac{\log_4(\nums)}{\lseq}$ that relates the total number and the length of the strands and can be interpreted as the penalty the channel induces by the permutation effect.

\section{Index-Based Concatenated Coding Scheme}

We propose an index-based concatenated coding scheme consisting of three codes: an  outer spatially-coupled low-density parity-check  (SC-LDPC) code, providing overall protection to the data and against non-drawn strands, an index code, which counteracts the loss of ordering in the presence of IDS errors, and an inner code whose main goal is to maintain synchronization.

The information $\bfu \in \field _{\chq} ^{\dimo}$ is encoded by an  $[\leno,\dimo,\frac{\nums}{2},m]_{\chq}$ SC-LDPC code  of output length $\leno$, input length $\dimo$, coupling length $\frac{\nums}{2}$, and coupling memory $\scmem$ over the field $\field_{\chq}$ \cite{felstrom1999time, lentmaier_SCLDPC_2010}. Note that there is a one-to-one mapping of the field $\field_{4}$ to the  DNA alphabet $\Sigma_4 = \{\mathsf{A},\mathsf{C},\mathsf{G},\mathsf{T}\}$. We interpret it as a field to allow linear operations over vectors in the DNA alphabet.  The resulting codeword $\bfw = (w_1,\ldots,w_{\leno}) \in \field_{\chq} ^{\leno}$ is split into $\nums$ equal-length data blocks as $\bfw = (\bfwi{1}, \ldots , \bfwi{\nums})$ such that $\bfwi{i} = (w^{(i)}_1,\ldots,w^{(i)}_{\bleno}) \in \field_{\chq}^{\bleno}$,   $\bleno = \frac{\leno}{\nums}$. Each index $i$ is encoded by an $[\nix,\kix]_\chq$ index code over $\field_{\chq}$ of even output length $\nix$ and even input length $\kix \geq \log_\chq (\nums)$ to an index codeword. Independently, the data block $\bfwi{i}$ is encoded by an  $[\nin,\kin]_\chq$ inner code over $\field_\chq$ of output length $\nin$, input length $\kin$, and rate $\Rin=\frac{\kin}{\nin}$, generating an inner data codeword of length $\frac{\bleno}{\Rin}$. We consider the marker-repeat (MR) codes presented in \cite{Srinivasavaradhan2021TrellisBMA, inoue_adaptivemarker2012} for the inner code, where the $\kin$-th input symbol is repeated once such that $\nin = \kin + 1$ and $\Rin=1-\frac{1}{\nin}$. 
Due to our decoding technique, we have higher error protection at the beginning and end of the DNA strands.  Thus, we split the index codeword in half and insert it at the beginning and the end of the inner codeword of the respective data block.  Finally, a random offset sequence is added to the generated strand resulting in the final encoded output strand $\bfx_i$, of length $\lseq =  \bleno \Rin^{-1} + \nix $.  The random offset is known to the decoder and supports the synchronization capability of the index and inner decoder and ensures that the nucleotides of the stored DNA strands are uniformly distributed over $\Sigma_4$. 
The final codeword list is then described by $\bfX = (\bfx_1,\dots,\bfx_\nums)$.
The overall code rate is   $\Rtot = 2 \cdot \Rout \frac{\bleno}{\lseq}$ in bits/nucleotide, with individual rates $\Rout = \frac{\dimo}{\leno}$, $\Rix = \frac{\kix}{\nix}$, and recall $\Rin = \frac{\kin}{\nin}$. 

\begin{figure}[t]
	\centering
	\begin{tikzpicture}
    \def\xxdist{5em};
	\def\yydist{3.0em};
	\def\bwidth{3.55em}

	\tikzstyle{seqblock} = [rectangle, draw, text centered, minimum height=1.6em,fill=black!5!, minimum width=\bwidth, outer sep=0]
    \tikzstyle{seqixblock} = [rectangle, draw, text centered, minimum height=1.6em,fill=black!1!, minimum width=0.2*\bwidth, outer sep=0]
			\tikzstyle{coding} = [rectangle, draw, text centered, rounded corners, minimum height=1.5em,fill=black!2!, minimum width=0.5*\bwidth]

		\node (u) at (0,0) {$\bfu$};
		\node[coding] (outenc) at (0.8*\xxdist, 0) {\bf Outer Enc.};
		\node[seqblock] (w1) at (2*\xxdist,0) {$\bfwi{1}$};
		\node[seqblock] (w2) at (2*\xxdist+\bwidth,0) {$\bfwi{2}$};
		\node[seqblock] (w3) at (2*\xxdist+2*\bwidth,0) {$\dots$};
		\node[seqblock] (w4) at (2*\xxdist+3*\bwidth,0) {$\bfwi{\nums}$};

        \node[seqblock] (ixseq1) at (2*\xxdist,-1*\yydist) {$\bfwi{1}$};
        \node[seqixblock, anchor=center] (ixseq2) at ($(ixseq1.west)+(-0.175*\bwidth,0)$) {\small ix};
        \node[seqixblock, anchor=west] (ixseq3) at ($(ixseq1.east)+(-0.00*\bwidth,0)$) {\small ix};
        \node (ix3) at (2*\xxdist+1.5*\bwidth,-1*\yydist) {$\dots$};

        \node[seqblock] (ixseq4) at (2*\xxdist+3*\bwidth,-1*\yydist) {$\bfwi{\nums}$};
        \node[seqixblock, anchor=center] (ixseq5) at ($(ixseq4.west)+(-0.175*\bwidth,0)$) {\small ix};
        \node[seqixblock, anchor=west] (ixseq6) at ($(ixseq4.east)+(-0.00*\bwidth,0)$) {\small ix};

		\node[coding, text width=5em] (enc1) at (2*\xxdist,-2*\yydist) {\bf Index/Inner Encoder};
        \node (enc3) at (2*\xxdist+1.5*\bwidth,-2*\yydist) {$\dots$};
        \node[coding, text width=5em] (enc4) at (2*\xxdist+3*\bwidth,-2*\yydist) {\bf Index/Inner Encoder};
		
		\node[coding, text width=12em] (channel) at (2*\xxdist+1.5*\bwidth,-3.4*\yydist) {\bf Multi-Draw IDS Channel};
		
		\node[coding, text width=8em] (indec) at (2*\xxdist+1.5*\bwidth,-4.8*\yydist) {\bf Joint Index and Inner Decoder};
		
		\node[coding] (outdec) at (0.8*\xxdist, -4.8*\yydist) {\bf Outer Dec.};
		
		\node (uhat) at (0.0, -4.8*\yydist) {$\hat{\bfu}$};
		
		\draw[->] (u.east) -- (outenc.west);
		\draw[->] (outenc.east) -- (w1.west);
		
        \draw[->] (w1.south) -- (ixseq1.north);
		\draw[->] (w4.south) -- (ixseq4.north);
		
        \draw[->] (ixseq1.south) -- (enc1.north);
		\draw[->] (ixseq4.south) -- (enc4.north);
		
		\draw[-] (enc1.south) -- ($(enc1.south)+(0,-0.5*\yydist)$) node[midway, left] {$\bfx_1$};
		\draw[-] (enc4.south) -- ($(enc4.south)+(0,-0.5*\yydist)$) node[midway, left] {$\bfx_\nums$};
		
		\draw[->] ($(enc1.south)+(0,-0.5*\yydist)$) -- ($(channel.north)+(-0.3*\xxdist,0)$);
		\draw[->] ($(enc4.south)+(0,-0.5*\yydist)$) -- ($(channel.north)+(0.3*\xxdist,0)$);
		
		\draw[-] ($(channel.south)+(-0.8*\xxdist,0)$) -- ($(channel.south)+(-0.8*\xxdist,-0.5*\yydist)$)  node[midway, left] {$\bfy_1$};
		\draw[-] ($(channel.south)+(-0.4*\xxdist,0)$) -- ($(channel.south)+(-0.4*\xxdist,-0.5*\yydist)$)  node[midway, left] {$\bfy_2$};
		\draw[-] ($(channel.south)+(-0.0*\xxdist,0)$) -- ($(channel.south)+(-0.0*\xxdist,-0.5*\yydist)$)  node[midway, left] {$\bfy_3$};
		\draw[draw=white] ($(channel.south)+(0.4*\xxdist,0)$) -- ($(channel.south)+(0.4*\xxdist,-0.5*\yydist)$)  node[midway, xshift=-0.7em] {$\dots$}; 
		\draw[-] ($(channel.south)+(0.8*\xxdist,0)$) -- ($(channel.south)+(0.8*\xxdist,-0.5*\yydist)$)  node[midway, left] {$\bfy_\totdraw$};
		
		\draw[->] ($(channel.south)+(-0.8*\xxdist,-0.5*\yydist)$) -- ($(indec.north)+(-0.4*\xxdist,0)$);
		\draw[->] ($(channel.south)+(-0.4*\xxdist,-0.5*\yydist)$) -- ($(indec.north)+(-0.2*\xxdist,0)$);
		\draw[->] ($(channel.south)+(-0.0*\xxdist,-0.5*\yydist)$) -- ($(indec.north)+(-0.0*\xxdist,0)$);
		\draw[->] ($(channel.south)+(0.8*\xxdist,-0.5*\yydist)$) -- ($(indec.north)+(0.2*\xxdist,0)$);
		
		\draw[->] (indec.west) -- (outdec.east) node [midway,above] {$q(w_t|\outY)$};
		
		\draw[->] (outdec.west) -- (uhat.east);
		
		\draw [decorate,decoration = {brace}] ($(w1.north west)+(0,0.05*\yydist)$) --  ($(w4.north east)+(0,0.05*\yydist)$) node [midway, above,yshift=0.05*\yydist] {$\bfw$};

\end{tikzpicture}
 \vspace{-2ex}
	\caption{Concatenated coding scheme for the communication over the multi-draw IDS channel. The index is split into two parts and is inserted at the beginning and the end of the block $\bfwi{i}$. The term $q(w_t|\outY)$ denotes the mismatched metric computed by the joint index-inner decoder.} 
	\label{fig:coding-scheme}
 \vspace{-3ex}
\end{figure}
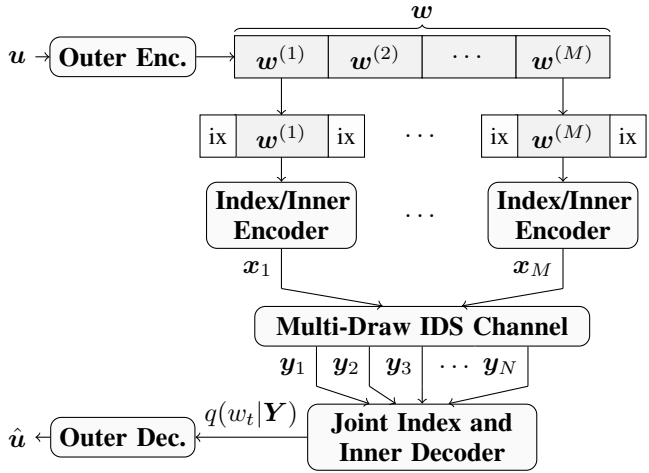

For the decoding procedure we introduce the following equivalent interpretation of the index and inner encoding as a joint process. We transform the integer $i$ to a vector $\bfii{i} \in \field_\chq^{\kix}$. This vector is split in half and placed at the beginning and end of the vector $\bfwi{i}$ resulting in the vector $\bfvi{i}=(v^{(i)}_1,\ldots,v^{(i)}_{\kix+\bleno})$, of length $\kix+\bleno$. Moreover, the  $[\nix, \kix]_\chq$ index code is generated by a serial concatenation of two equal $[\frac{\nix}{2},\frac{\kix}{2}]_\chq$ codes. Subsequently, the vector $\bfvi{i}$ is encoded at the index positions by the $[\frac{\nix}{2},\frac{\kix}{2}]_\chq$ code and symbolwise by the inner MR code at the positions corresponding to $\bfwi{i}$.
The coding/decoding scheme is depicted in Fig.~\ref{fig:coding-scheme}. %

\section{Joint Index and Inner Decoding}\label{sec:decoding}

The channel introduces two main impairments that need to be combated: the loss of ordering of the DNA strands due to the permutation effect, and the possibility that a strand is not drawn due to the drawing nature. However, the latter also provides inherent redundancy due to possible multiple copies of the same DNA strand that can be leveraged. Moreover, at the receiver side, the loss of synchronization within each strand due to insertion and deletion events is challenging. In general, we follow a \emph{mismatched decoding} approach due to the decoder's channel uncertainty and complexity constraints. Instead of considering all DNA strands in the received \listset\ $\outY$ jointly to produce an estimate of the information, $\hat{\bfu}$, we propose the following sub-optimal decoding scheme.

First, we infer symbolwise APPs using a joint index and inner MAP decoder by means of the BCJR algorithm for each received strand $\bfy_1,\dots,\bfy_\totdraw$ independently. Optionally, clustering of the DNA strands using the APPs is performed. In that way the clustering makes use of the coding gain provided by the index and inner code.
For combining the APPs within a cluster, the APPs are symbolwise multiplied, generating the mismatched APPs for one cluster. For a given cluster, by making hard-decisions on the index APPs, we estimate the block index $\hat{i}$ within $\bfw$. Therefore, assuming optimal clustering, we can actually benefit from the multi-copy gain of the channel for the index estimation. Once more, the APPs of the data block belonging to the same index $\hat{i}$ are multiplied, generating the mismatched APPs $q(w_t^{(i)}|\outY)$. The overall ordered APPs $q(w_t|\outY)$ are then passed to the outer decoder, which neglects any possible correlations of the symbolwise APPs. The outer code is specifically designed to handle \emph{block-fading} effects, as described in Section~\ref{sec:outercode}. Combining APPs after inner and before outer decoding by symbolwise multiplication is inspired by the well performing \emph{separate decoding} strategy of \cite{Maarouf2022ConcatenatedCF}. %

\subsection{Inner MAP Decoding}

We consider the case of decoding a single output strand $\bfy$ given an input $\bfv$ by means of the BCJR decoder over the joint code trellis of the index and inner code. We drop the superscript of the sequence $\bfv$ due to the permutation effect of the channel.  We follow the concept introduced in \cite{davey_reliable_2001}, refined in \cite{briffa_improved_2010}. For a detailed view, we refer to our prior work \cite{Maarouf2022ConcatenatedCF}. 

The IDS errors introduce a synchronization loss that destroys the Markov property of the joint index-inner code. However, we can form a hidden Markov model by introducing the drift $d_t$, defined as the number of insertions minus the number of deletions that occurred at time $t$, as a hidden state variable. The essence of the drift states is that the sequence $d_0,\dots,d_{\bleno+\kix}$ forms a Markov chain and the outputs after time $t+1$ are independent of the previous states, restoring the Markov property. Hence, we can formulate forward and backward recursions for a BCJR decoder, whose final output APPs are 
$p(v_t | \bfy ) = \frac{p(v_t,\bfy)}{p(\bfy)} \propto \sum_{d,d'} p(v_t,\bfy,d,d')$,
where $d$ and $d'$ denote realizations of the RVs $d_t$ and $d_{t+1}$, respectively.

\subsection{Clustering}

Common approaches perform clustering on the received strands before error-correction decoding using approximation techniques for the Levenshtein distance (minimum number of IDS operations), e.g., %
\cite{antkowiak_low_2020, rashtchian_clustering_2017}. Here, we present a clustering approach of complexity $\cO(\totdraw^2 \lseq)$ that leverages the coding gain of the index and inner code.%

Given the APPs $p(v_t|\bfy_j)$, for each strand $\bfy_j$ a vector $\bfrj{j} =(r^{(j)}_1,\ldots,r^{(j)}_{\binL}) \in \mathbb{R}^{\binL}$ of binary log-likelihood ratios (LLRs) is obtained using marginalization, where $\mathbb{R}$ denotes the reals and $\binL$ is the length of the corresponding binary vector of $\bfv$. %
The clustering decision is made by evaluating  the pairwise Euclidean distance 
\begin{equation*}
   \textstyle{ \eucd{i,j} = \sqrt{\sum_{\ell=1}^{\binL} \left(r^{(i)}_\ell - r^{(j)}_\ell\right)^2}}
\end{equation*}
for any two vectors $\bfrj{i}$ and $\bfrj{j}$, $i \neq j$.  
Let $D_\mathrm{E}^{\mathrm{intra}}$ be the RV of the Euclidean distance between two received sequences $\bfrj{i}$ and $\bfrj{j}$ originating from the same stored DNA strand. Assuming no correlations between the binary LLRs, $D_\mathrm{E}^{\mathrm{intra}}$ converges asymptotically in $\binL$ to a normal distribution $\mathcal{N}(\mu_\mathrm{D},\sigma^2_\mathrm{D})$ with mean $\mu_\mathrm{D}$ and variance $\sigma^2_\mathrm{D}$ \cite{dawkins_theoretical_2021}. By sampling realizations of $D_\mathrm{E}^{\mathrm{intra}}$, it can be observed that this is  a good approximation for $\binL > 100$ for fitted parameters $\mu_\mathrm{D}$ and $\sigma^2_\mathrm{D}$. Consequently, given the computed pairwise distances $\eucd{i,j}$ for $i,j \in \{1,\dots,\totdraw\}$, $i \neq j$, strands are considered to stem from the same DNA strand if 
$\eucd{i,j} \leq \mu_\mathrm{D} + \omega \cdot \sigma_\mathrm{D}$,
where $\omega \in \mathbb{R}^{+}$ is a design parameter and $\mathbb{R}^{+}$ denotes the set of positive real numbers.

Let the $k$-th cluster, $1 \leq k \leq \numc$, be formed as $\outC_k \triangleq (\bfy_{k,1}, \dots, \bfy_{k,|\outC_k|})$, where $\bfy_{k,j}$ is the $j$-th strand placed in cluster $\outC_k$ and $|\outC_k|\neq 0$. 
Denote the overall clustering output as the \listset\ $\outC = (\outC_1,\dots,\outC_{\numc})$,  with $1 \leq \numc \leq \totdraw $ and $\sum_{k=1}^{\numc} |\outC_k| = \totdraw$. For each cluster $k$, we multiply the respective APPs giving the mismatched rule
\begin{align*}
    q(v_t|\outC_k) \approxprop \prod _{j=1} ^{|\outC_k|} \frac{p(v_t|\bfy_{k,j})}{p(v_t)^{|\outC_k|-1}}\,.
\end{align*}

\subsection{Index and Multiple DNA Strand Decoding}

The index code and recovering the index information of each strand/cluster at the receiver side is the key point of tackling the permutation effect and leveraging the multi-copy gain of the channel. For each cluster $\outC_k$, we obtain the soft information of the indices by extracting 
\begin{align*}
  \bfq(\bfii{i}|\outC_k) \triangleq &\left(q(v_1|\outC_k),{\tiny \dots},q(v_{\kix/2}|\outC_k), \right.  \\
  & \qquad \left. q(v_{\bleno+{\kix/2}+1}|\outC_k),\dots, q(_{\bleno+\kix}|\outC_k)\right)\,.   
\end{align*}
We perform hard-decisions on $\bfq(\bfii{i}|\outC_k)$ to compute an estimate index $\hat{i}_{k}$ for every cluster $1 \leq k \leq \numc$, which will determine how the strand/cluster is grouped. Let $\outS_i$ be the set of clusters with decision on index $i$, where $0 \leq |\outS_i| \leq \numc$. 
For all data positions of the $i$-th block, i.e., for symbols of $\bfwi{i}$, we compute the  APPs according to the mismatched rule
\begin{align*}
    q(w^{(i)}_t|\outY) \approxprop \prod _{k \in \outS_i} \frac{q(w^{(i)}_t|\outC_k)}{q(w^{(i)}_t)^{|\outS_i|-1}}\,,
\end{align*}
where $q(w^{(i)}_t|\outY) = \frac{1}{\outq}$ when $\outS_i = \emptyset$.

As a final step, according to the position of block $\bfwi{i}$ in the sequence $\bfw$, the mismatched APPs are given to the outer decoder, which outputs an estimate $\hat{\bfu}$.

\section{Achievable Information Rates}\label{sec:air}
    \vspace{-0.5ex}
We compute AIRs of mismatched decoders for fixed index and inner codes. Specifically, we compute i.u.d. \emph{BCJR-once} rates ($\Rbcjr$), measured in bits/nucleotide, which are defined as the symbolwise mutual information between the input and its corresponding LLRs with  uniform input distribution \cite{Kavcic2003DE, muller_capacity_2004, soriaga_determining_2007}. %
The BCJR-once rates serve as an appropriate measure of an AIR for an outer decoder that ignores possible correlations between the symbolwise estimates $q(w_t|\outY)$ given by the inner MAP decoder. Moreover, by using a mismatched decoding metric $q(\bfw|\outY)$ instead of the true metric $p(\bfw|\outY)$, the rate $\Rbcjr$ only decreases. Consider the mismatched-decoder  decoding metric 
\begin{align*}
    q(\bfw|\outY) = \prod_{i=1}^{\nums} \prod_{t=1}^{\bleno} q(\wit|\outY)\,.
\end{align*}
The mutual information $\mi (\bfw;\outY)$ can be bounded from below as (for clarity of presentation, we do not distinguish between RVs and their realizations in our notation) 
\begin{align*}
    \mi(\bfw;\outY) &\geq %
     \mathbb{E}_{\bfD} \left[ \sum_{i,t} \mathbb{E}_{\wit,\outY | \bfD} \left[ \log_2 \frac{q(\wit|\outY)}{p(\wit)}\right]\right]\,,
\end{align*}
where $\mathbb{E}_{X}[\cdot]$ denotes expectation with respect to the RV $X$. 

We define the LLR representation
\begin{align*}
    \llr_{i,t}(a) = \sum _{k \in \outS_{i}}  \sum_{j=1} ^{|\outC_k|} \ln \frac{q(w^{(i)}_t=a|\bfy_{k,j})}{q(w^{(i)}_t=0|\bfy_{k,j})}\,,
\end{align*}
where $\bfy_{k,j}$ denotes the $j$-th strand of $\outC_k$.
 Similar to \cite{Kavcic2003DE}, we use the mismatched LLR representation to formally define
\begin{align*}
&\Rbcjr \triangleq \\
&\qquad\mathbb{E}_{\bfD} \left[  \lim_{\nums \lseq \to \infty} \frac{1}{\nums \lseq} \sum_{i=1}^\nums \sum_{t=1} ^{\bleno} \mi \left( w^{(i)}_t; \llr_{i,t}(w^{(i)}_t )\right) \right]\,,%
\end{align*}
which depends on the multi-draw parameters $\beta$ and $c$, IDS channel parameters $\pri$, $\prd$, and $\prs$, and is evaluated for a fixed index and inner coding/decoding scheme. Assuming ergodicity for a single strand, we follow a simulation-based approach to calculate $\Rbcjr$. We simulate for long strand lengths $\widetilde{\lseq} = 100\,000$ (correspondingly $\widetilde{\bleno}=(\widetilde{\lseq}-\nix)\Rin$ and $\widetilde{\nums}=4^{\beta \widetilde{\lseq}}$), reflecting $\nums \lseq \to \infty$ for a fixed $\beta$, and approximate $\mathbb{E}_{\bfD} \left[\cdot \right]$ by the Monte-Carlo method. 
Hence, by averaging over a large number of draw realizations, $\Phi$, we approximate $\Rbcjr$ as 
\begin{align*}
    &\Rbcjr \approx \\
    &\frac{1}{\Phi} \sum_{\phi=1}^\Phi \frac{1}{\widetilde{\nums} \widetilde{\lseq}} \sum_{i=1}^{\widetilde{\nums}} \sum_{t=1}^{\widetilde{\bleno}} \left(\log_2 \outq + \log_2 \frac{{\mathrm e}^{\llr_{i,t}(\wit)}}{\sum_{a \in \field_{\outq}}{\mathrm e}^{\llr_{i,t}(a)}} \right)\,.
\end{align*}

\section{Outer Code}\label{sec:outercode}

Due to its drawing effect, the DNA storage channel as seen by the outer code resembles a block-fading channel when considering a finite number of strands $\nums$ \cite{weinberger_dna_2022}. Hence, it also shares its afflictions, most importantly its non-ergodic property. In particular, an \emph{outage} event occurs when not enough strands are drawn for a specific block. Formally, an outage event occurs when the instantaneous mutual information between the input and output of the channel is lower than the transmission rate $\Rout$. %
We consider the BCJR-once version of the information-outage probability adapted from \cite{buckingham_informationoutage_2008}, to which we refer as the \emph{mismatched information-outage probability} $\qout$. It incorporates our mismatched decoding approach, the fixed i.u.d. input, and the dispersion due to the finite blocklength phenomena. Let $\rbcjr$ denote the instantaneous BCJR-once information density for the finite length regime and fixed draw realization $\bfd$, computed as
\begin{align*}
\rbcjr = \frac{1}{\nums \lseq} \sum_{i=1}^\nums \sum_{t=1} ^{\bleno} \mi \left( w^{(i)}_t; \llr_{i,t}(w^{(i)}_t )\right)\,.
\end{align*}
Then, $\qout$ is formally defined as
\begin{align*}
    \qout = {\rm Pr}\left( \rbcjr < \Rout' \right) \geq \pout\,,
\end{align*}
where $\Rout'=2\Rout$ such that the outer code rate is measured in bits/nucleotide and $\pout$ is the true outage probability. $\qout$ gives a lower bound on the FER for a given encoder and mismatched decoder pair, i.e., using the mismatched  metric described Section~\ref{sec:air}, for fixed finite number of strands, fixed finite blocklength, and fixed channel parameters. We approximate $\qout$ by the Monte-Carlo method for a given $\Rout'$.

The diversity order of the outer code is an important parameter that influences the slope of the FER curve for low IDS probabilities. By coding over blocks of $\bfw$, the outer code can provide a \emph{diversity} gain. %
We consider protograph-based SC-LDPC codes for the outer code, as they  achieve high diversity for the block-fading channel \cite{hassan_diversityscldpc_2014}. %
 We optimize the protograph following the procedure  in  \cite{hassan_diversityscldpc_2014} based on  the density evolution outage (DEO) probability bound.
The optimization works by first finding a block LDPC protograph with DEO close to a target $\qout = 10^{-3}$ based on the joint index-inner code. %
Then, an optimization over the edge spreading to create the SC-LDPC protograph is performed. In order to design a non-binary code, random
edge weights from $\field_4$ are assigned to the edges of the protograph throughout the optimization. In the  optimization, we fix the coupling length to $\frac{\nums}{2}$ and the coupling memory to $m=2$. %

\section{AIR and FER Results}

We analyze our proposed coding scheme for the multi-draw IDS channel by means of AIRs, information-outage probability for the outer code, and FER simulations for different coverage depths $c$ and fixed $\beta$. We set the IDS channel parameters to $\pri=0.017$, $\prd=0.020$, and $\prs=0.022$, motivated by the results of the experiment from \cite{Srinivasavaradhan2021TrellisBMA}. Additionally, we present FER results on their experimental data. Moreover,  we fix the inner code rate to $\Rin = \frac{10}{11}$ and perform full-window belief propagation decoding with a maximum of $100$ iterations for the SC-LDPC outer code. The index codes $[3,1]_4$ (AIR simulations) and $[6,2]_4$ (information-outage/FER simulations), both with $\Rix = \frac{1}{3}$, are obtained via an exhaustive graph search algorithm optimizing the code's Levenshtein distance spectrum (see Table~\ref{tab:indexcode}) \cite{Sewell1998clique2, Maarouf2022ConcatenatedCF}.
\begin{table}[t]
\begin{center}
\caption{Index Codes} \label{tab:indexcode}
\vspace{-1.8ex}
\setlength{\tabcolsep}{1pt}
\begin{tabular}{c c c l l l l }

\toprule 
   Code & & & \multicolumn{4}{l}{Codebook} \\
         \midrule
    $[3,1]_4$ & & & \multicolumn{4}{l}{$(0,3,3)$, $(1,0,2)$, $(2,1,1)$, $(3,2,0)$}  \\
      \midrule
      $[6,2]_4$ & & & $(0,0,1,0,1,1)$, & $(0,0,3,3,2,2)$, & $(0,2,2,2,2,0)$, & $(1,1,1,0,2,3)$, \\
      & & & $(1,1,2,2,0,1)$, & $(1,1,3,3,3,0)$, & $(1,3,0,3,1,1)$, & $(2,0,1,1,0,0)$, \\
      & & & $(2,0,3,2,3,3)$, & $(2,2,2,0,3,2)$, & $(2,2,3,1,1,1)$, & $(2,3,0,2,2,2)$, \\
      & & & $(3,1,0,1,3,3)$, & $(3,1,1,2,1,2)$, & $(3,3,2,1,1,0)$, & $(3,3,3,0,0,3)$ \\
         \bottomrule
\end{tabular}
\end{center}
\vspace{-4.5ex}
\end{table}

\begin{figure}[t]
    \centering
    
    \begin{tikzpicture}
	\begin{axis}[
	width = 0.97\columnwidth,
	xmin=1,   xmax=10,
	ymin=0.5,	ymax=2,
	xticklabel style = {/pgf/number format/fixed, /pgf/number format/precision=6},
	xtick={1,2,3,4,5,6,7,8,9,10},
	grid = both,
    grid style={gray,opacity=0.5,dotted},
	legend cell align={left},
	legend style={font=\footnotesize,at={(1.0,0.0)},anchor=south east},
	xlabel = {$c$},
	ylabel = {$\Rbcjr$ {\footnotesize[bits/nucleotide]}},
  xlabel style={
	yshift=0.7ex,
	name=label},
 ylabel style={
	xshift=0.0ex,
    yshift=-2.0ex,
	name=label}
	]

	\addplot+ [color=red,mark=*,mark options={line width = 0.5pt, fill=white}] table [col sep=comma, x=c, y=I] {Figures/air/m16_L100000_mr10-11/SimAI_data=0_N=1454560_q=4_M=16_Npay=100001_Nind=2_code=MR-11_off=1_betaL=2_idxc=rep-1_ixpos=s_gidx=1_clus=none.csv};
	\addlegendentry{$\Rix=1$ with an ordering genie};

	\addplot+ [color=orange,mark=square*,mark options={solid, line width = 0.5pt, fill=white}, solid] table [col sep=comma, x=c, y=I] {Figures/air/m16_L100000_mr10-11/SimAI_data=0_N=1454560_q=4_M=16_Npay=100001_Nind=2_code=MR-11_off=1_betaL=2_idxc=rep-1_ixpos=s_gidx=0_clus=none.csv};
	\addlegendentry{$\Rix=1$}

	{\addplot+ [color=teal,mark=diamond*, solid,mark options={solid, line width = 0.5pt, fill=white}] table [col sep=comma, x=c, y=I] {Figures/air/m16_L100000_mr10-11/SimAI_data=0_N=1454560_q=4_M=16_Npay=100001_Nind=6_code=MR-11_off=1_betaL=6_idxc=lev_ixpos=s_gidx=0_clus=none.csv};}
	{\addlegendentry{$\Rix=\nicefrac{1}{3}$}}

	{\addplot+ [color=blue,mark=triangle*, solid,mark options={solid, line width = 0.5pt, fill=white}] table [col sep=comma, x=c, y=I] {Figures/air/m16_L100000_mr10-11/SimAI_data=0_N=1454560_q=4_M=16_Npay=100001_Nind=6_code=MR-11_off=1_betaL=6_idxc=lev_ixpos=s_gidx=0_clus=euclid_full.csv};}
	{\addlegendentry{$\Rix=\nicefrac{1}{3}$ with clustering};}

    \addplot+[mark=none, black] table [col sep=comma] 
	{
        1,1.1722059598916335
	    2,1.587630224905016
		5,1.8074785003907383
		10,1.8181146706421552
		};
	{\addlegendentry{Noiseless}}

     \addplot[mark=none, color=gray, dashdotted] coordinates {(1,1.8181487609316194) (10,1.8181487609316194)};

    \end{axis}
    
    \end{tikzpicture}
     \vspace{-2ex}
        \caption{
     BCJR-once rates versus coverage depth $c$ for different index coding rates $\Rix$ and decoding techniques. Fixed parameters are $\beta = 2\cdot 10^{-5}$, $\pri=0.017$, $\prd=0.020$, $\prs= 0.022$, and $\Rin = \frac{10}{11}$. For clustering, we determined $D_\mathrm{E}^{\mathrm{intra}} \sim \cN(938, 6.4^2)$ and use $\omega = 5$. The dash-dotted gray line represents the rate limit due to the inner code rate.
 }
    \label{fig:air-m16-mr67}
     \vspace{-4.5ex}
\end{figure}
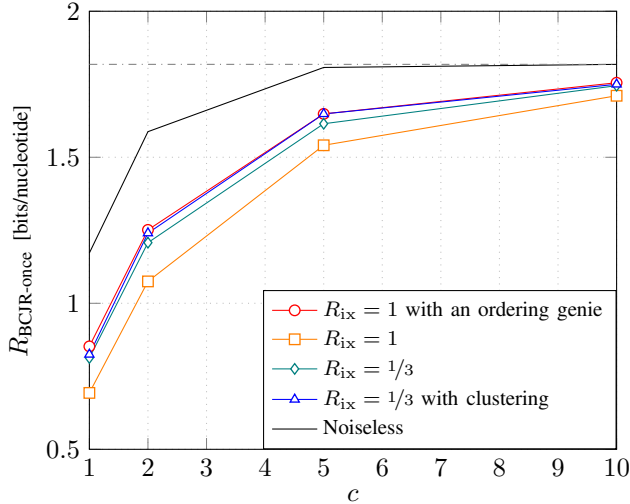

Fig.~\ref{fig:air-m16-mr67} shows the AIR $\Rbcjr$ of different coding/decoding schemes for $\beta = 2\cdot10^{-5}$. We observe the unavoidable rate loss due to the non-drawing strand and permutation effect for low coverage depth $c$ which, however, diminishes for increasing values of $c$ due to the multi-copy gain. For an overall benchmark, we include the AIR of an optimal index-based scheme over the noiseless channel (i.e., only drawing and permuting effects). With IDS noise, we  include a benchmark for an index-based coding approach given an index genie in the decoding process, i.e., we artificially exclude the permutation loss of the channel. The rate gap to the noiseless scenario can be  explained by the IDS noise and is also due to our chosen sub-optimal coding and mismatched decoding approach. Moreover, protecting the index with a strong code seems crucial since in the non-coded case received strands may be grouped incorrectly. For the given parameters, the designed index code performs very close to  the genie ordering curve, while the non-coded index curve suffers from a big rate loss. Further, applying our clustering method in combination with a coded index attains the index genie benchmark since the clustering enhances the quality of the index decisions.

\begin{figure}[t]
    \centering
\input{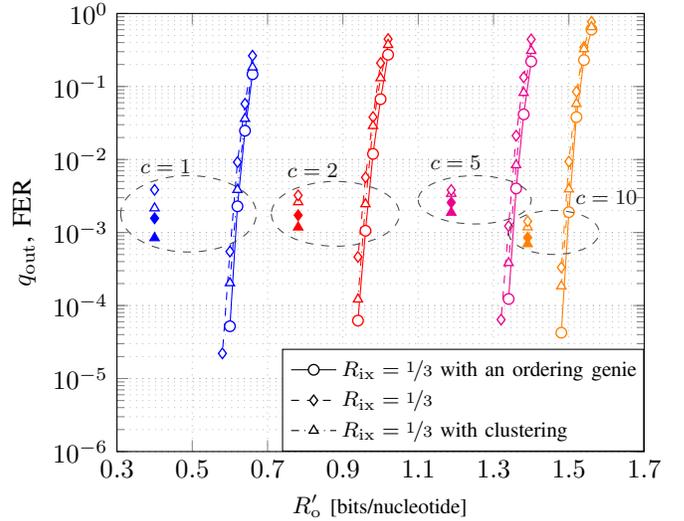}
     \vspace{-4.5ex}    
    \caption{Information-outage probability $\qout$ and FER versus outer code rate $\Rout'=2\Rout$, for fixed $\leno = 23040$, $\nums = 256$, $L = 110$, $\beta \approx 0.36$, $\pri=0.017$, $\prd=0.020$, $\prs= 0.022$, and $\Rin = 0.9183 \approx \frac{10}{11}$. For clustering, we determined $D_\mathrm{E}^{\mathrm{intra}} \sim \cN(31.52, 5.44^2)$ and use $\omega = 2.5$. The different curves represent $\qout$ for different coding/decoding schemes.  The stand-alone markers correspond to the FER performance when using an optimized SC-LDPC outer code on synthetic (solid shaped) and experimental (empty shaped) data.}
    \label{fig:Outage_Prob}
    \vspace{-2.3ex}
\end{figure}

Fig.~\ref{fig:Outage_Prob} shows the information-outage probability $\qout$ versus the outer code rate $\Rout'=2\Rout$ for different coverage depths $c$ and decoding approaches
for $\nums=256$ input strands and fixed index code of rate $\Rix=\frac{1}{3}$. 
For a fixed $\Rout'$, the corresponding $\qout$ serves as a lower bound for any code's FER with that rate and length, and a decoder following our mismatched decoding rule.
In general, we see a similar behavior of our proposed coding/decoding schemes as for the AIRs. In the same figure, we show the FER performance of a protograph-based SC-LDPC outer code (stand-alone solid shaped markers). The SC-LDPC protograph is optimized  individually for each considered coding/decoding scheme and then randomly lifted and assigned random edge weights from $\field_4$. %
We observe an expected rate loss that can be explained by the limited protograph search space. The corresponding optimized (block) LDPC photographs are shown in Table~\ref{tab:DE_protographs}. Finally, we also include FER results on experimental data from \cite{Srinivasavaradhan2021TrellisBMA} (stand-alone empty shaped markers). Notably, the penalty in performance, due to the fact that the experimental channel may suffer from other additional noise impairments, is small. In addition, the SC-LDPC protographs are optimized on synthetic samples and not on experimental data which also contributes to the loss in performance.
\begin{table}[t]
\begin{center}
\caption{Block LDPC  Photographs for Different Values of $c$}%
\label{tab:DE_protographs}
\vspace{-2.3ex}
\begin{tabular}{c c c c }
\toprule 
   $c$ & \multicolumn{2}{c}{Protograph$^a$} & $\Rout'$ \\[0.5ex]
    & Coded-index & Coded-index + clustering &\\
      \midrule
         $1$ & $\left(\begin{smallmatrix} 1 & 1 & 1 & 0 & 1 \\ 1 & 1 & 1 & 1 & 0 \\ 1 & 0 & 1 & 0 & 1 \\ 0 & 1 & 0 & 2 & 0\end{smallmatrix}\right)$ & $\left(\begin{smallmatrix} 1 & 1 & 1 & 0 & 1 \\ 0 & 0 & 2 & 2 & 0 \\ 2 & 0 & 0 & 0 & 1 \\ 0 & 2 & 0 & 1 & 0\end{smallmatrix}\right)$ & 0.3750\\[0.4 cm]%
         $2$ & $\left(\begin{smallmatrix} 1 & 2 & 0 & 0 & 1 \\ 0 & 0 & 1 & 1 & 1 \\ 2 & 1 & 2 & 1 & 0 \end{smallmatrix}\right)$ & $\left(\begin{smallmatrix} 1 & 1 & 0 & 1 & 0 \\ 1 & 0 & 2 & 1 & 1 \\ 1 & 2 & 1 & 0 & 1 \end{smallmatrix}\right)$ & 0.7812\\[0.4cm]%
         $5$ & $\left(\begin{smallmatrix} 2 & 1 & 0 & 2 & 2\\ 0 & 2 & 2 & 1 & 1 \end{smallmatrix}\right)$ & $\left(\begin{smallmatrix} 2 & 1 & 0 & 2 & 2\\ 0 & 2 & 2 & 1 & 1 \end{smallmatrix}\right)$ & 1.1876\\[0.3 cm] %
         $10$ & $\left(\begin{smallmatrix} 1 & 2 & 1 & 1 & 0 & 0 & 0 & 1 & 1 & 2\\ 1 & 0 & 2 & 0 & 2 & 1 & 2 & 0 & 1 & 0 \\ 0 & 1 & 0 & 1 & 1 & 2 & 0 & 2 & 1 & 1 \end{smallmatrix}\right)$  & $\left(\begin{smallmatrix} 1 & 1 & 1 & 1 & 0 & 0 & 0 & 1 & 2 & 1\\ 1 & 1 & 0 & 0 & 2 & 1 & 2 & 2 & 0 & 1 \\ 0 & 1 & 2 & 1 & 1 & 2 & 0 & 0 & 1 & 1 \end{smallmatrix}\right)$ & 1.3906\\\bottomrule%
\end{tabular}
\end{center}
   \footnotesize{$^a$Optimized edge spreading is performed on the given protographs, with a coupling length of $128$ and a coupling memory of $m = 2$. Recall $\Rout'=2\Rout$.}
    \vspace{-6.0ex}
    \end{table}

\section{Conclusion}

We proposed a practical index-based concatenated coding scheme for the multi-draw IDS channel, a model approximating the DNA storage process. Further, we presented low-complexity decoding techniques for this setup. Our AIR results show that, in the presence of IDS errors, protecting the index with a strong code is crucial. Finally, we proposed explicit code constructions with FER performance of around  $10^{-3}$  for synthetic and experimental data and $256$ input strands.

\clearpage

\balance

\bibliographystyle{IEEEtran}
\bibliography{refs}

\end{document}